\documentclass[aps,prl,reprint,twocolumn,superscriptaddress]{revtex4-2}

\usepackage{tikz}
\usepackage{amsmath,amssymb}
\usepackage{hyperref}
\usepackage{braket}

\begin{document}

\title{Logarithmic corrections to black hole entropy from minimum-assumptions discretization}

\author{Pelayo V. Calzada}
\affiliation{Departamento de Física, Universidad de Alicante, Campus de San Vicente del Raspeig, E-03690 Alicante, Spain}
\affiliation{Institut für Physik, Humboldt-Universität zu Berlin, Zum Großen Windkanal 6, 12489 Berlin, Germany}
\author{Ana Alonso-Serrano}
\affiliation{Institut für Physik, Humboldt-Universität zu Berlin, Zum Großen Windkanal 6, 12489 Berlin, Germany}
\affiliation{Max-Planck-Institut für Gravitationsphysik (Albert-Einstein-Institut),
Am Mühlenberg 1, 14476 Potsdam, Germany}
\author{Ernesto Contreras}
\affiliation{Departamento de Física, Universidad de Alicante, Campus de San Vicente del Raspeig, E-03690 Alicante, Spain}
\author{Pedro Bargueño}
\affiliation{Departamento de Física, Universidad de Alicante, Campus de San Vicente del Raspeig, E-03690 Alicante, Spain}
\date{\today}

\begin{abstract}
We introduce here a general  model, under agnostic and minimum assumptions, to uniquely find the entropy area law with a corresponding fixed logarithmic correction term. In this approach, the horizon is discretized into generic Planck-scale cells representing coarse-grained indistinguishable geometric structures, and it is just the statistical combinatorial counting that determines the form of the entropy terms. We highlight the role of the assumptions and their comparison with known models providing fixed logarithmic contributions to the entropy.

\end{abstract}

\maketitle


\emph{Introduction.—} The relation between the entropy of a black hole and the area of its horizon, \(S = k_B \frac{A}{4 \ell_p^2}\), plays a central role in our understanding of the interplay between quantum theory and gravitation \cite{Bekenstein1973,Hawking1975}. This result has not only provided a key testing ground for candidate theories of quantum gravity, but has also inspired derivations of the gravitational field equations from thermodynamic arguments \cite{Jacobson1995}, as well as alternative viewpoints in which gravity itself is interpreted in thermodynamic terms (see, e.g., \cite{Padmanabhan2010,Padmanabhan2010bis,Bargueno2022complexity}).
\\
\\
Despite the success of this semiclassical picture, the interpretation of this entropy, and in particular for microscopic models, 
a clear identification of the microscopic degrees of freedom responsible for black hole entropy, together with a universally accepted counting scheme, remains elusive. Progress has been made in several directions, including string theory \cite{STROMINGER1996}, loop quantum gravity (LQG) \cite{Kaul2000,Meissner2004}, conformal field theory techniques \cite{Carlip2000}, and approaches inspired by Wheeler’s “it from bit” paradigm \cite{Wheeler1989,Makela2019,Davidson2019,Bargueno2022,Bargueno2023}. Nevertheless, there is still no consensus on the fundamental nature of these degrees of freedom.
\\
\\
A broadly shared expectation is that quantum effects modify the entropy beyond the leading area term by introducing subleading corrections. In particular, a logarithmic contribution of the form \(S = k_B\left(\frac{A}{4 \ell_p^2} + c \ln \frac{A}{\ell_p^2}\right)\) is commonly found, where the coefficient \(c\) depends on the underlying framework \cite{Carlip2000,Kaul2000,Alonso-Serrano:2020dcz}. For instance, LQG yields \(c=-1/2\) in a \(U(1)\) treatment and \(c=-3/2\) in the full \(SU(2)\) formulation (see \cite{Perez2017} for a review), while conformal methods reproduce the latter value \cite{Carlip2000}. In contrast, string-theoretic calculations typically lead to different coefficients \cite{Banerjee2011}. It is worth mentioning here that other interpretations of this entropy, such as an entanglement entropy  \cite{Bombelli:1986rw,Srednicki:1993im}, also predict the same modified form of the subleading terms \cite{Solodukhin:2011gn}.
\\
\\
More recently, it has been shown that the Bekenstein--Hawking entropy can arise under very general assumptions—namely, the existence of a UV completion and the validity of the semiclassical Euclidean path integral—without invoking specific microscopic models such as string theory or holography \cite{Bala2024,Bala2024bis}. Although these results strongly support the universality of the area law, they do not address the structure of subleading corrections.
\\
\\
In this work, we propose a minimal statistical framework that not only reproduces the leading area dependence but also naturally generates a logarithmic correction with coefficient \(c = -1/2\). Our construction does not rely on a detailed microscopic theory of quantum gravity, in line with \cite{Bala2024,Bala2024bis}, and instead points toward a universal origin for both the leading and subleading contributions. After specifying the basic assumptions of the model, we explicitly compute the number of microstates and the associated entropy, showing that the expected structure emerges without the need for fine tuning.
\\
\\
 \emph{A minimal statistical model.—} Our first assumption is that, following Wheeler's approach ~\cite{Wheeler1989}, there are some holographic degrees of freedom corresponding to some bits of information encoded in each
 Planck area on the horizon. Therefore, we model the horizon as discretized into \(N\) fundamental cells of area \(\ell_p^2\), where \(N = A/\ell_p^2\). 
  Following the emergent interpretation pushed forward by Padmanabhan ~\cite{Padmanabhan2010bis}, these $N$ {\it spacetime atoms} ~\cite{Padmanabhan2010,Padma2010tris} count the microscopic degrees of freedom and, even more, they satisfy a holographic equipartition ~\cite{Padma2010tris} which provides a direct link between the macroscopic and microscopic descriptions. This assumption is connected to the standard consideration of a minimum resolution for the constituents of the black hole area. Let us note as well the relation of this approach with previous studies on thermodynamics of Geodesic Local Causal Diamonds from minimum area \cite{Alonso-Serrano:2021eju}. 
\\
\\
Our second assumption is that we represent each degree of freedom by a position $x$ and a momentum $p$ in a two-dimensional phase space, so that the elementary phase-space cell has volume $h^2$, with $h$ the Planck constant. Rather than conventional particles, these elements should be interpreted as coarse-grained labels for the geometric degrees of freedom that emerge from the underlying microscopic structure. No additional dynamical information or structure is assigned to these effective degrees of freedom. By the principle of maximum entropy, all accessible microstates then carry equal statistical weight, and the absence of additional structure renders all configurations indistinguishable. This indistinguishability introduces a Gibbs factor $1/N!$ in the microstate count, which is the origin of the logarithmic correction to the entropy in our model.


Finally, the phase space per degree of freedom is taken to have a finite volume proportional to the area $A$. We implement this by restricting the momentum to a finite interval $[-\Lambda,\Lambda]$, so that the phase space becomes $A \times [-\Lambda,\Lambda]$. This restriction is another key ingredient of our framework, as it encodes a fundamental ultraviolet cutoff, a generic feature expected in quantum gravity. Such a cutoff arises naturally in several contexts, including minimal length scenarios \cite{Garay:1994en,Kempf1995}, polymer quantization \cite{Ashtekar2003}, and noncommutative geometry \cite{Doplicher1995,Snyder1947}. We parametrize it as $\Lambda \equiv \frac{\alpha \, h}{2 \, \ell_p}$, where $\alpha$ is a dimensionless number of order unity and the factor of two is introduced for later convenience.
\\
\\
\emph{Microstate counting.—} The number of accessible microstates then follows from integrating over this phase space,

\begin{equation}\label{eq:omega-integral}
\Omega_{N} = \frac{1}{N!\; h^{2 N}} \int_{A^{N}} d^{2 N}x \int_{[-\Lambda,\Lambda]^{2 N}} d^{2 N}p .
\end{equation}

Since the integrals factorize, we compute:
\begin{equation}
\int_{A^{N}} d^{2 N}x = A^{N}, \qquad \int_{[-\Lambda,\Lambda]^{2N}} d^{2N}p = (2\Lambda)^{2N}.
\end{equation}
Then, after performing the integrations, the number of microstates can be rewritten in terms of an effective elementary area in configuration space defined as
\begin{equation}
L^2 := \left(\frac{h}{2\Lambda}\right)^2,
\end{equation}
and becomes
\begin{equation}\label{omega}
\Omega_{N} = \frac{1}{N!} \left( \frac{A}{L^2} \right)^{N} =   \alpha^{2N} \frac{N^N}{N!}.
\end{equation}

Once we have constructed the number of microstates, the thermodynamic entropy is straightforwardly obtained as ($k_{B}=1$)
\begin{equation}
S = \ln \Omega_{N}  =  2N \ln \alpha +  \ln \frac{N^{N}}{N!}.
\end{equation}

After including the full expansion for the $N!$ we get
\begin{widetext}

\begin{equation}
    S =N(1+2\ln\alpha)
-\frac12\ln N
-\sum_{k=1}^{\infty}
\frac{B_{2k}}{2k(2k-1)N^{2k-1}}
-\frac12\ln(2\pi),
\end{equation}
where $B_{2k}$ are the Bernoulli numbers.
    
\end{widetext}

The parameter $\alpha$ encodes the relationship between the ultraviolet cutoff $\Lambda$ and the Planck scale,
and is fixed by the single consistency condition that the leading term reproduces the Bekenstein-Hawking
area law. Imposing $N(1 + 2\ln\alpha) = N/4$ yields
\begin{equation}
    \alpha = e^{-3/8} \approx 0.687,
\end{equation}
which is indeed of order one, as required. With this value, the entropy reads
\begin{widetext}
\begin{equation}
  S =
\frac{A}{4l_p^2}
-\frac12 \ln\!\left(\frac{A}{l_p^2}\right)
-\sum_{k=1}^{\infty}
\frac{B_{2k}}{2k(2k-1)}
\left(\frac{l_p^2}{A}\right)^{2k-1}-\frac12 \ln(2\pi).
\end{equation}
\end{widetext}



\emph{Discussion.—} A word is in order regarding the statistical assumptions underlying our model. Three ingredients enter the description of the microstates.
\\
\\
First, all accessible microstates carry equal statistical weight, equivalent to a constant Hamiltonian throughout phase space. This is the natural choice for a model explicitly agnostic about the microscopic dynamics of the horizon degrees of freedom. Since our goal is to count geometric configurations, rather than to describe their time evolution, the relevant question is not which Hamiltonian governs the system, but how many arrangements are compatible with a given horizon area. In this sense, our framework is better understood as a combinatorial or microcanonical model: the count in Eq.~(\ref{omega}) is the fundamental object for calculating the entropy, not a statement about thermal equilibrium at a particular temperature. This is consistent with the general philosophy of effective models in quantum gravity, where the statistical properties of the horizon are expected to emerge from counting arguments rather than from a specific dynamical theory.
\\
\\
Second, the label description of each degree of freedom carries no intrinsic physical meaning, so configurations differing only by a permutation are assigned to the same physical macrostate. Treating the degrees of freedom as indistinguishable can be interpreted to reflect the nature of fundamental particles, as formalized in relativistic quantum field theory. It also aligns with LQG models of black hole entropy \cite{Ghosh:2013iwa}. Still, it may not be an essential hypothesis to obtain logarithmic corrections to the black hole entropy from a minimal statistical framework. We are currently working on such an alternative model, built from distinguishable degrees of freedom, to deepen the understanding of the minimum general assumptions generating corrections of this kind.
\\
\\
Third, the phase space of each degree of freedom is taken proportional to the horizon area. Since the area is the only physical information the model is built on, a minimal-assumptions approach suggests relating the phase space to it through a single functional dependence. Linear proportionality can be considered the simplest such choice, and it yields a microstate count consistent with the Bekenstein-Hawking entropy. This has a discrete analogue in LQG \cite{Perez2017}, where the area is built from individual puncture contributions and the states available to each puncture are controlled by its area quantum, scaling with the total area.
\\
\\
A crucial observation is that the parameter $\alpha$, although fixed by the consistency condition on the leading term, plays no role in determining the coefficient of the logarithmic correction. Indeed, the term $-\frac{1}{2}\ln N$ in the entropy arises entirely from the Stirling expansion of $\ln(N^N/N!)$ and is therefore independent of $\alpha$. Within the specific assumptions of our minimal model -- in particular, the complete indistinguishability of all horizon degrees of freedom -- the coefficient $c = -1/2$ follows directly from the Stirling expansion. This suggests that complete indistinguishability leads to a universal prediction for this coefficient. Deviations from this value, such as the $c = -3/2$ found in the SU(2) formulation of LQG \cite{Perez2017} or the $c = -1/2$ in the U(1) formulation, would signal the presence of additional structure (e.g., multiple spin labels, nontrivial gauge groups, or partial distinguishability). Thus, while our model provides a minimal benchmark, the logarithmic correction may ultimately serve as a discriminant between different quantum gravity frameworks.
\\
\\
One of the most interesting features of the present model is precisely the emergence of the logarithmic correction with coefficient $-1/2$. This result is obtained from a minimal label description of the horizon discretization. No detailed microscopic realization is assumed; the degrees of freedom of the model are interpreted as effective building blocks emerging from the underlying microscopic structure. The logarithmic correction arises from the Stirling expansion of the factorial term in the microstate count. Therefore, its origin is purely combinatorial and does not depend on dynamical details. This suggests that the correction may be universal, reflecting general properties of discrete geometric degrees of freedom rather than specific features of a particular quantum gravity model. This is particularly interesting in light of previous results obtained in conformal field theory \cite{Carlip2000} and LQG \cite{Perez2017}, where the same coefficient appears across different settings. By contrast, the leading area term carries no comparable universality: its coefficient is fixed by matching against a cutoff, and is therefore subject to the corresponding ambiguities of scale. Moreover, higher-order corrections are found to be suppressed in inverse powers of the area \cite{Carlip2000, Kaul2000}. In our approach, the same characteristics emerge from a minimal statistical model, with higher-order corrections arising naturally from the Bernoulli expansion. These terms are also suppressed by inverse powers of the area and therefore become relevant only for Planck-scale black holes, providing further evidence that the entropy subleading terms may have a universal origin. 
\\
\\
It is worth noting that the same features -- a UV cutoff, an area-law scaling, and logarithmic corrections -- also appear in the entanglement entropy of a massless scalar field across a spherical entangling surface in four-dimensional flat spacetime \cite{Solodukhin:2011gn}. In that context, the area term scales as $A/\epsilon^2$, where $\epsilon$ is a UV cutoff, and the logarithmic correction carries a coefficient that depends on the matter content. In our model, the area term is $A/(4\ell_p^2)$ after fixing the dimensionless constant $\alpha$, which effectively sets the scale of the cutoff $\Lambda$. Upon identifying the UV cutoff $\epsilon$ in the entanglement entropy calculation with the Planck scale $\ell_p$, both frameworks yield an area-law entropy with a logarithmic correction whose coefficient is a pure universal number. 
This parallelism suggests as well that this universal behavior of the correction from microstates could also be connected with the entanglement entropy and allows for the future development of a model from minimum assumptions which encapsulates the entropy encoding quantum correlations. A detailed investigation of this correspondence lies beyond the scope of this letter but represents a promising direction for future work.

\vspace{0.5cm}
\begin{acknowledgments}
PVC acknowledges support from Generalitat Valenciana Grants No. CIACIF/2021/268 and No. CIBEFP/2026/145. AA-S is funded by the Deutsche Forschungs-
gemeinschaft (DFG, German Research Foundation) — Project ID 516730869. PB and EC acknowledge financial support from the Generalitat Valenciana through PROMETEO PROJECT CIPROM/2022/13 and from MCIN/AEI through Project PID2025-171322NB-C21. 
EC is funded by the Beatriz Galindo contract BG23/00163 
(Spain). AA-S also acknowledges partial support through Grant No. PID2023-149018NB-C44 (funded by MCIN/AEI/10.13039/501100011033). We thank JA Miralles for valuable suggestions and a critical reading of a first version of the manuscript.
\end{acknowledgments}

\bibliographystyle{apsrev4-2}
\bibliography{refs}

@article{Bekenstein1973,
  author = {Bekenstein, J. D.},
  title = {Black holes and entropy},
  journal = {Phys. Rev. D},
  volume = {7},
  pages = {2333--2346},
  year = {1973}
}

@article{Hawking1975,
  author = {Hawking, S. W.},
  title = {Particle Creation by Black Holes},
  journal = {Commun. Math. Phys.},
  volume = {43},
  pages = {199--220},
  year = {1975}
}

@article{Carlip2000,
  author = {Carlip, S.},
  title = {Logarithmic corrections to black hole entropy from the Cardy formula},
  journal = {Class. Quant. Grav.},
  volume = {17},
  pages = {4175--4186},
  year = {2000}
}

@article{Jacobson1995,
  title = {Thermodynamics of Spacetime: The Einstein Equation of State},
  author = {Jacobson, Ted},
  journal = {Phys. Rev. Lett.},
  volume = {75},
  issue = {7},
  pages = {1260--1263},
  numpages = {0},
  year = {1995},
  month = {Aug},
  publisher = {American Physical Society},
  doi = {10.1103/PhysRevLett.75.1260},
  url = {https://link.aps.org/doi/10.1103/PhysRevLett.75.1260}
}

@article{Padmanabhan2010,
  title = {Surface density of spacetime degrees of freedom from equipartition law in theories of gravity},
  author = {Padmanabhan, T.},
  journal = {Phys. Rev. D},
  volume = {81},
  issue = {12},
  pages = {124040},
  numpages = {12},
  year = {2010},
  month = {Jun},
  publisher = {American Physical Society},
  doi = {10.1103/PhysRevD.81.124040},
  url = {https://link.aps.org/doi/10.1103/PhysRevD.81.124040}
}

@article{Padmanabhan2010bis,
doi = {10.1088/0034-4885/73/4/046901},
url = {https://dx.doi.org/10.1088/0034-4885/73/4/046901},
year = {2010},
month = {mar},
publisher = {},
volume = {73},
number = {4},
pages = {046901},
author = {Padmanabhan, T},
title = {Thermodynamical aspects of gravity: new insights},
journal = {Reports on Progress in Physics}
}

@article{Kaul2000,
  title = {Logarithmic Correction to the Bekenstein-Hawking Entropy},
  author = {Kaul, Romesh K. and Majumdar, Parthasarathi},
  journal = {Phys. Rev. Lett.},
  volume = {84},
  issue = {23},
  pages = {5255--5257},
  numpages = {0},
  year = {2000},
  month = {Jun},
  publisher = {American Physical Society},
  doi = {10.1103/PhysRevLett.84.5255},
  url = {https://link.aps.org/doi/10.1103/PhysRevLett.84.5255}
}

@article{Meissner2004,
doi = {10.1088/0264-9381/21/22/015},
url = {https://dx.doi.org/10.1088/0264-9381/21/22/015},
year = {2004},
month = {oct},
publisher = {},
volume = {21},
number = {22},
pages = {5245},
author = {Krzysztof A Meissner},
title = {Black-hole entropy in loop quantum gravity},
journal = {Classical and Quantum Gravity}
}

@article{Banerjee2011,
title ={Logarithmic corrections to and black hole entropy: a one loop test of quantum gravity},
journal = {JHEP},
volume = {143},
pages = {(2011)},
author = {S. Banerjee and R. K. Gupta and I. Mandal and A. Sen},
}

@article{STROMINGER1996,
title = {Microscopic origin of the Bekenstein-Hawking entropy},
journal = {Physics Letters B},
volume = {379},
number = {1},
pages = {99-104},
year = {1996},
issn = {0370-2693},
doi = {https://doi.org/10.1016/0370-2693(96)00345-0},
url = {https://www.sciencedirect.com/science/article/pii/0370269396003450},
author = {Andrew Strominger and Cumrun Vafa}
}

@article{Wheeler1989,
title={proceedings of the 3rd International Symposium
 on Foundations of Quantum Mechanics (Physical Society of
 Japan (1990), Tokyo, 1989)},
journal = {Proceedings of the 3rd International Symposium
 on Foundations of Quantum Mechanics},
pages = {354},
year = {1990},
author = {J. Wheeler},
}

@article{Bargueno2023,
  title = {Threefold way to black hole entropy},
  author = {Bargue\~no, Pedro and Contreras, Ernesto},
  journal = {Phys. Rev. D},
  volume = {107},
  issue = {6},
  pages = {066013},
  numpages = {6},
  year = {2023},
  month = {Mar},
  publisher = {American Physical Society},
  doi = {10.1103/PhysRevD.107.066013},
  url = {https://link.aps.org/doi/10.1103/PhysRevD.107.066013}
}

@article{Bargueno2022,
  title = {Minimal model for the Bekenstein-Hawking entropy},
  author = {Bargue\~no, Pedro and Contreras, Ernesto},
  journal = {Phys. Rev. D},
  volume = {106},
  issue = {6},
  pages = {066001},
  numpages = {7},
  year = {2022},
  month = {Sep},
  publisher = {American Physical Society},
  doi = {10.1103/PhysRevD.106.066001},
  url = {https://link.aps.org/doi/10.1103/PhysRevD.106.066001}
}

@article{Makela2019,
author = {M\"{a}kel\"{a}, Jarmo},
title = {Wheeler’s it from bit proposal in loop quantum gravity},
journal = {International Journal of Modern Physics D},
volume = {28},
number = {10},
pages = {1950129},
year = {2019},
doi = {10.1142/S0218271819501293},
}

@article{Davidson2019,
  title = {From Planck area to graph theory: Topologically distinct black hole microstates},
  author = {Davidson, Aharon},
  journal = {Phys. Rev. D},
  volume = {100},
  issue = {8},
  pages = {081502},
  numpages = {7},
  year = {2019},
  month = {Oct},
  publisher = {American Physical Society},
  doi = {10.1103/PhysRevD.100.081502},
  url = {https://link.aps.org/doi/10.1103/PhysRevD.100.081502}
}

@article{Perez2017,
doi = {10.1088/1361-6633/aa7e14},
url = {https://dx.doi.org/10.1088/1361-6633/aa7e14},
year = {2017},
month = {oct},
publisher = {IOP Publishing},
volume = {80},
number = {12},
pages = {126901},
author = {Perez, Alejandro},
title = {Black holes in loop quantum gravity},
journal = {Reports on Progress in Physics}
}

@article{Bala2024,
  title = {Microscopic Origin of the Entropy of Astrophysical Black Holes},
  author = {Balasubramanian, Vijay and Lawrence, Albion and Mag\'an, Javier M. and Sasieta, Martin},
  journal = {Phys. Rev. Lett.},
  volume = {132},
  issue = {14},
  pages = {141501},
  numpages = {6},
  year = {2024},
  month = {Apr},
  publisher = {American Physical Society},
  doi = {10.1103/PhysRevLett.132.141501},
  url = {https://link.aps.org/doi/10.1103/PhysRevLett.132.141501}
}

@article{Bala2024bis,
  title = {Microscopic Origin of the Entropy of Black Holes in General Relativity},
  author = {Balasubramanian, Vijay and Lawrence, Albion and Mag\'an, Javier M. and Sasieta, Martin},
  journal = {Phys. Rev. X},
  volume = {14},
  issue = {1},
  pages = {011024},
  numpages = {31},
  year = {2024},
  month = {Feb},
  publisher = {American Physical Society},
  doi = {10.1103/PhysRevX.14.011024},
  url = {https://link.aps.org/doi/10.1103/PhysRevX.14.011024}
}

@article{Padma2010tris,
author = {Padmanabhan, T.},
title = {EQUIPARTITION OF ENERGY IN THE HORIZON DEGREES OF FREEDOM AND THE EMERGENCE OF GRAVITY},
journal = {Modern Physics Letters A},
volume = {25},
number = {14},
pages = {1129-1136},
year = {2010},
doi = {10.1142/S021773231003313X}
}

@article{Kempf1995,
author = {Kempf, A. and Mangano, G. and Mann, R. B.},
title = {Hilbert space representation of the minimal length uncertainty relation},
journal = {Phys. Rev. D},
volume = {52},
pages = {1108},
year = {1995}
}

@article{Ashtekar2003,
author = {Ashtekar, A. and Fairhurst, S. and Willis, J. L.},
title = {Quantum gravity, shadow states, and quantum mechanics},
journal = {Class. Quant. Grav.},
volume = {20},
pages = {1031},
year = {2003}
}

@article{Doplicher1995,
author = {Doplicher, S. and Fredenhagen, K. and Roberts, J. E.},
title = {The quantum structure of spacetime at the Planck scale and quantum fields},
journal = {Commun. Math. Phys.},
volume = {172},
pages = {187},
year = {1995}
}

@article{Snyder1947,
author = {Snyder, H. S.},
title = {Quantized space-time},
journal = {Phys. Rev.},
volume = {71},
pages = {38},
year = {1947}
}

@article{Solodukhin:2011gn,
    author = "Solodukhin, Sergey N.",
    title = "{Entanglement entropy of black holes}",
    eprint = "1104.3712",
    archivePrefix = "arXiv",
    primaryClass = "hep-th",
    doi = "10.12942/lrr-2011-8",
    journal = "Living Rev. Rel.",
    volume = "14",
    pages = "8",
    year = "2011"
}

@article{Garay:1994en,
    author = "Garay, Luis J.",
    title = "{Quantum gravity and minimum length}",
    eprint = "gr-qc/9403008",
    archivePrefix = "arXiv",
    reportNumber = "IMPERIAL-TP-93-94-20",
    doi = "10.1142/S0217751X95000085",
    journal = "Int. J. Mod. Phys. A",
    volume = "10",
    pages = "145--166",
    year = "1995"
}

@article{Alonso-Serrano:2020dcz,
    author = "Alonso-Serrano, Ana and Li{\v{s}}ka, Marek",
    title = "{Quantum phenomenological gravitational dynamics: A general view from thermodynamics of spacetime}",
    eprint = "2009.03826",
    archivePrefix = "arXiv",
    primaryClass = "gr-qc",
    doi = "10.1007/JHEP12(2020)196",
    journal = "JHEP",
    volume = "12",
    pages = "196",
    year = "2020"
}

@article{Alonso-Serrano:2021eju,
    author = "Alonso-Serrano, Ana and Li{\v{s}}ka, Marek",
    title = "{Thermodynamics of spacetime from minimal area}",
    eprint = "2107.08749",
    archivePrefix = "arXiv",
    primaryClass = "gr-qc",
    doi = "10.1103/PhysRevD.104.084043",
    journal = "Phys. Rev. D",
    volume = "104",
    number = "8",
    pages = "084043",
    year = "2021"
}

@article{Bombelli:1986rw,
    author = "Bombelli, Luca and Koul, Rabinder K. and Lee, Joohan and Sorkin, Rafael D.",
    title = "{A Quantum Source of Entropy for Black Holes}",
    reportNumber = "PRINT-86-0371 (SYRACUSE)",
    doi = "10.1103/PhysRevD.34.373",
    journal = "Phys. Rev. D",
    volume = "34",
    pages = "373--383",
    year = "1986"
}

@article{Srednicki:1993im,
    author = "Srednicki, Mark",
    title = "{Entropy and area}",
    eprint = "hep-th/9303048",
    archivePrefix = "arXiv",
    reportNumber = "LBL-33754, CFPA-93-02",
    doi = "10.1103/PhysRevLett.71.666",
    journal = "Phys. Rev. Lett.",
    volume = "71",
    pages = "666--669",
    year = "1993"
}

@article{Ghosh:2013iwa,
    author = "Ghosh, Amit and Noui, Karim and Perez, Alejandro",
    title = "{Statistics, holography, and black hole entropy in loop quantum gravity}",
    eprint = "1309.4563",
    archivePrefix = "arXiv",
    primaryClass = "gr-qc",
    doi = "10.1103/PhysRevD.89.084069",
    journal = "Phys. Rev. D",
    volume = "89",
    number = "8",
    pages = "084069",
    year = "2014"
}

@article{Bargueno2022complexity,
  author  = {Bargue\~no, P. and Fuenmayor, E. and Contreras, E.},
  title   = {Complexity factor for black holes in the framework of the {N}ewman--{P}enrose formalism},
  journal = {Annals of Physics},
  volume  = {443},
  pages   = {169012},
  year    = {2022},
  doi     = {10.1016/j.aop.2022.169012}
}

\end{document}